# Charge tunable structural phase transitions in few-layer tellurium: a step toward building mono-elemental devices

Cong Wang[1], Jingsi Qiao[1], Yuhao Pan[1], Linwei Zhou[1], Xianghua Kong[1], Zhihai Cheng[1], Yang Chai[2] and Wei Ji[1, *]

[1]*Beijing Key Laboratory of Optoelectronic Functional Materials & Micro-Nano Devices, Department of Physics, Renmin University of China, Beijing 100872, P. R. China*

[2]*Department of Applied Physics, The Hong Kong Polytechnic University, Hung Hom, Kowloon, Hong Kong, P. R. China*

\* wji@ruc.edu.cn

Recently, a covalent-like quasi-bonding was unveiled for the inter-chain interaction in a promising semiconductor, few-layer Tellurium. Such quasi-bond offers comparable bond lengths and strengths with those of a typical Te-Te covalent bond, which may lead to much easier transformations between the quasi-bonds and the covalent bonds. Here, we show a few structural phase-transitions among four Te allotropes in few-layer Te under charge doping. In particular, the semiconducting $α$-phase can transform into a smaller-bandgap $β$-phase, even-smaller-bandgap $γ$-phase and then metallic $δ$-phase in a Te bilayer. In a tri-layer, a metallic chiral $α+δ$ phase is more stable under initial electron doping, leading to the appearance of chirality. Variations of electronic structures aside, these transitions are accompanied by the emergence or elimination of inversion centers ($α$-$β$, $α$-$γ$, $α$-$α+δ$), structural anisotropy ($α$-$γ$, $γ$-$δ$) and chirality ($α$-$α+δ$), which could result in substantial changes in optical and other properties. In light of this, this work opens the possibility toward building mono-elemental electronic and optoelectronic heterostructures or devices. It also offers a platform for studying relations between charge doping and electric/optical properties.

Mono- and Few-layer Tellurium were recently unveiled a promising quasi-one-dimensional layered material by both experimental demonstrations[1-5] and theoretical predictions[6,7]. A distinctive feature of it lies in its rich number of allotropes that four phases were theoretically predicted[6,7]. The most-stable few-layer α-phase (Fig. 1a) is comprised of helical Te chains bonded through covalent-like quasi-bonding (CLQB)[7-16] across chains. It offers extremely high carrier mobility, with the bandgap varying from 1.17 eV (2L) to 0.30 eV (bulk). The bandgaps for 2L gradually reduce in the meta-stable β- (0.62 eV, Fig. 1b) and meta-stable γ-phases (0.26 eV, Fig. 1c), respectively. In contrast to these three phases, the strain-induced δ-phase is metallic and chain-like [7], which might be a promising candidate for atomic electric wires. These phases offer coexistence, competition and transitions among substantially different electronic structures.

Although these properties are rather striking, the β, γ and δ phases are, however, less stable in neutral and strain-free FL-Te and may spontaneously transfer into the most stable alpha phase. It is thus crucial to maintain these less-stable phases using a long-lasting and feasible way. Recently, the ionic liquid gating technique has been rapidly developing [17-23], which achieved structural transitions among three configurations in $SrCoO_{2.5}$ based oxides using $H^+$ or $O^{2-}$ ions[20]. If those ions are not involved in doping, sole charge doping can induce TMDs transforming from their semiconducting 2H phase to the metallic 1T or 1T' phase [17,21,24,25]. However, these transitions are resulted from changes of bond angles but are not from breakdown or formation of new bonds as required in the transitions among these Te allotropes. Few-layer Te (FL-Te) is a rather special case that the covalent characteristic of inter-chain CLQB largely lowers the energy costs for forming inter-chain and breaking intra-chain covalent bonds. It would be thus interesting to know if sole-charge-doping could manipulate covalent bonds in structural phase-transitions of Te and maintain those meta-stable Te phases, which may substantially extend the fields that doping techniques apply.

Fundamental interests[26-32] of controllable phase transitions in 2D materials aside, reversibly [17, 19, 20, 24, 33] and locally [18] controlled structural transitions among different Te phases may lead to diverse potential applications of Te as functional materials and devices, e.g. improved metal-semiconductor contracts[21, 22, 34-36], photonic devices[37], catalysts[38] and rewritable data storage[39]. Beyond this, an all-Te transistor, e.g. $α$, $β$ or $γ$ for channels and $δ$ for electrodes, might be realized by controlled phase changes, which achieves a mono-elemental transistor and is substantially different from previous transistors with metal or graphene electrodes. Along the lines of all-Te transistor, Te few-layers are also promising in building all-Te heterojunctions, all-Te quantum wells or all-Te solar cells.

Here, we theoretically predicted structural transitions among these Te phases solely induced by charge doping. Either 2L or 3L exhibits three transitions among four phases. These transitions are accompanied by elimination or emergence of structural inversion symmetry and chirality, gradually reduced bandgaps, likely topological states or structural isotropy, which suggest diverse potential applications in terms of electronics, optics and related fields. We revealed the reasoning of all these transitions at the wavefunction level, including the emergence or extinction of central inversion symmetry or chirality. We also showed Raman spectra and band alignments of $α$, $β$ and $γ$ phases and discussed electric properties of these phases.

## Results

**Phase diagram.** Fig. 1a-1d show the top- and side-views of $α$-, $β$-, $γ$- and $δ$-Te bilayers [6, 7, 40], respectively. The $β$-phase [space groups $P2/m$ (No. 10) for odd and $P2_1/m$ (No. 11) for even numbers] has one additional mirror/inversion symmetry to the $α$-phase [$P2$ (No.3) and $P2_1$ (No.4)]. The $γ$-phase [$P3/m_1$(No. 164)] shows a $C_{3v}$ rotation symmetry and is structurally isotropic, in different from the $α$ and $β$-phases. In neutral bilayers, $α$-Te is slightly more stable than γ-Te (1meV/Te) and β-Te is unstable tending to transform to α-Te according to phonon frequency calculations [40] (Supplementary Figure S1). As previously mentioned, δ-Te [$P2_1/m$ (No. 11)] could be transferred from α-Te by over 20% uniaxial tensile strain along the helical chain direction and could meta-stably exist after the strain releases [7]. In δ-Te, the helical chains forming α-Te are stretched into zigzag chains.

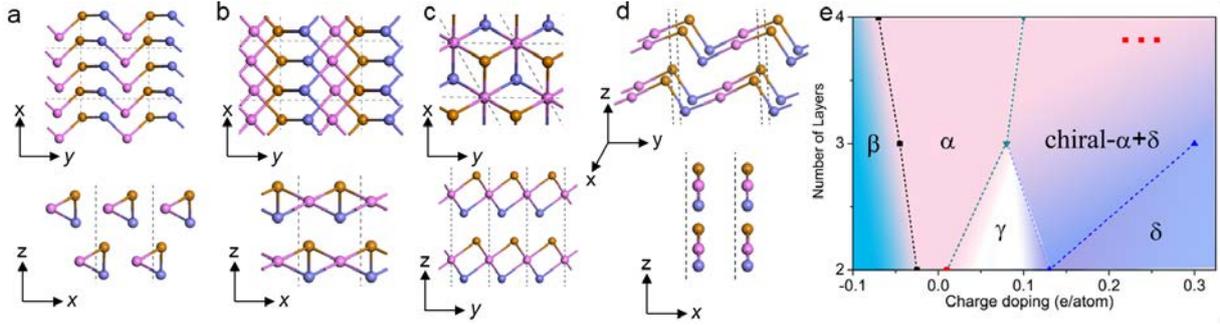

**Fig. 1** Schematic models of four phases of Te and aphase diagram of 2L-4L upon charge doping. **a-d** Top- and side-views of α- **a**, β- **b**, γ- **c** and δ-Te bilayers **d**. Orange, voilet and slate-blue balls represent Te atoms in different sublayer along the z direction. **e** Phase diagram of 2L-, 3L- and 4L-Te upon charge doping. Solid smybols are the exact transtion points and are connected by dotted lines showing bonddaries between different phases.

Fig. 1e shows a diagram of these phases from 2L to 4L under charge doping. For a bilayer, α-Te spontaneously transfers into β-Te under either hole (0.03 $h$/Te) or electron (0.06 $e$/Te) doping (Supplementary Figure S2). Here, 0.1 $e$/Te or 0.1 $h$/Te in α phase is equivalent to a carrier density of $\pm 2.4 \times 10^{14}$ $e$/cm$^2$ and is usually reachable in ionic liquid gating techniques. The electron doped β-Te is, however, energetically less favored in comparison with electron-doped γ-Te. At a doping level of 0.01 $e$/Te, γ-Te becomes energetically more stable than other phases until a level of 0.13 $e$/Te, beyond which δ-Te is the most stable one. In terms of 3L, the hole doped α-β transition was still found at 0.04 $h$/Te, however, the α-γ transition vanishes, instead, an α to chiral-α+δ transition occurs at a doping level of 0.08 $e$/Te. The absent of γ-trilayers is consistent with its rapidly reduced stability in 3L and thicker layers[7]. The chiral-α+δ phase is a mixed phase particularly found in 3L that those three layers show left-hand α (l-α), δ and right-hand *α* (*r-α*) chains, respectively. These layers may have ten different arrangements to form a 3L. Detailed discussion is available in Supplementary Table S1. We also found a transition from *α*-Te to meta-stable *l-α+r-α* chiral *α*-Te in bilayers, as discussed in Supplementary Figure S3. At a rather high doping level of 0.3 *e*/Te, the chiral-*α+δ*-trilayer fully transforms into a pure *δ*-phase. The mixed chiral structure is even more complicated in 4L that it contains 25 possible arrangements of *l-α*-Te, *r-α*-Te and *γ*-Te chains. Their energy differences between different arrangements become indecisive when the doping level beyond 0.10 *e*/Te. We therefore leave this area undefined in our present phase diagram and will discuss it elsewhere. In the following paragraphs, we intend to understand these three phase transitions, namely *α-β*, *γ-δ* and *α*-to-chiral-*α-δ*.

**Emergence of mirror symmetry.** Fig. 2a shows the top-view of both an *α*- and a *β*-bilayer in one panel. The vertical motion of the Te3 or Te4 atom plays a key role in the *α-β* transition. We monitored three Te-Te bond lengths under hole or electron doping and plotted them in Fig. 2b, respectively. They explicitly indicate that the transition involves an elongated Te2-Te4, a shortened Te2-Te3 and a nearly unchanged Te1-Te2 bond lengths by either a hole or electron doping. Here, Te3 and Te4 are imaging atoms of two adjacent chains and Te1 and Te5 are those of two adjacent unit cells within a chain. As mentioned earlier, the *β*-bilayer is energetically more stable than the *α*-bilayer in doping levels higher than 0.03 *h*/Te or 0.06 *e*/Te, respectively. Both phases coexist up to 0.05 *h*/Te or 0.10 *e*/Te, at which the *α*-bilayers structurally transform to *β*-bilayers without barriers. The transition regions were thus marked with meshes in Fig. 2b. As a result of the varied bond lengths, lattice constant *b* expands under electron doping but shrinks in the positively charged region (Supplementary Figure S4). The variation of *b* exceeds 0.1 Å compared between neutral and a doping level of 0.05 *h*/Te, indicating a potential application of Te few-layers in charge driven actuators.

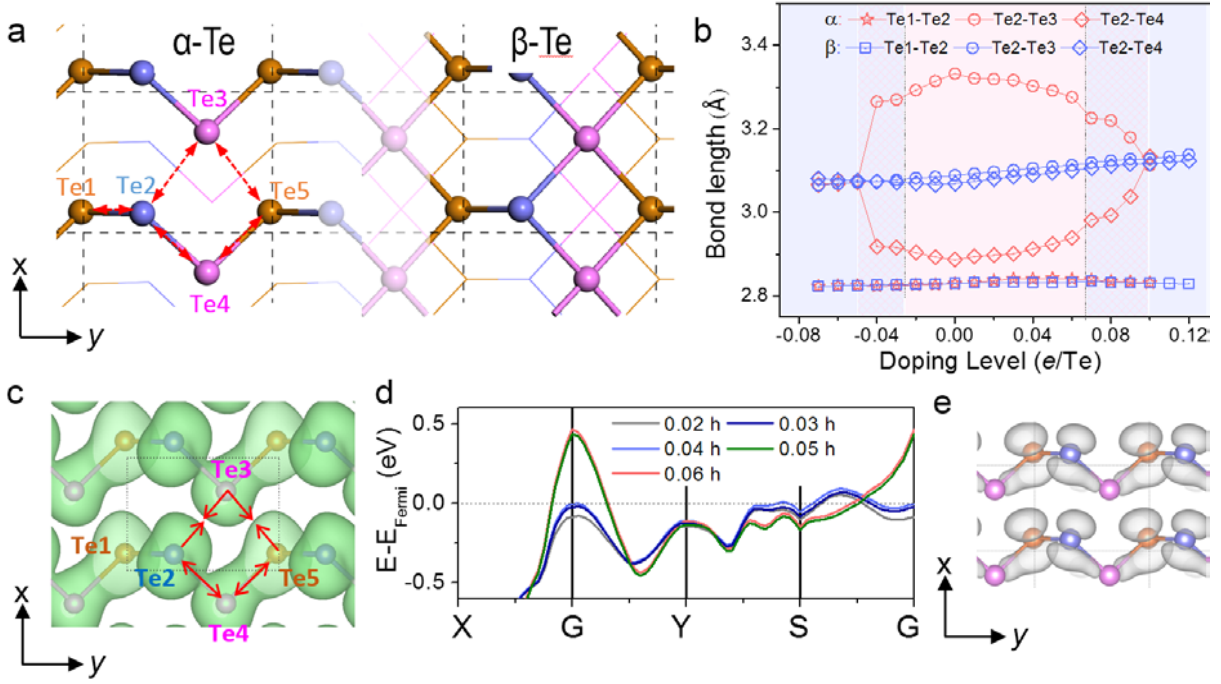

**Fig. 2** *α-β* phase tansition in bilayer Te. **a** Top view of *α*- and *β*-Te bilayers. Atoms in the top layer were represented by balls while thoese in the bottom layer were shown by lines. There are three Te atoms in a unitcell of a helical chain, which are mapped with different colors. **b** Te-Te bond lengths as a function of electron/hole doping level. Blue and pink regions represent the energeticlly favored regions of the *α* and *β* phases, respectively, while the transition regions were decorated with meshes. **c** Differential charge density of a hole doped *α*-Te bilayer (isosurface value of $0.0001 e/\text{Bohr}^3$). Here, green isosufacre coutours show charge reduction after hole doping. **d** Evolution of the highest valence band (VB1) under different hole doping levels. **e** Top view of the wavefunction norm of the VB1 state of *α* phase at G with an isosurface value of $0.001\ e/\text{Bohr}^3$.

Phonon frequency calculations suggest that electron-doped *β*-bilayers are unstable and tend to transform into chiral-*α*-bilayers. We thus focus on the hole-doping side and document the details of electron-doping in the Supplementary Figure S5. Fig. 2c plots a hole-doped differential charge densities (DCD) plotted between a 0.03 *h*/Te doped and a neutral *α*-bilayers. Charge reduction was found at the inter-atomic region between Te4 and Te2/Te5, suggesting weakened Te4-Te2/Te5 bonds. These weakened bonds are thus prone to move Te3/Te4 toward an adjacent chain; this transfers an *α*-bilayer to a *β*-bilayer. Such redistribution of charges could be well explained by doping-dependent bandstructures. At doping levels under 0.05 *h*/Te, eigenstates of the highest valence band (VB1) were unoccupied between the S and G points of the first BZ (Fig. 2d). These states are neither bonding nor anti-boding states (Supplementary Figure S6). Therefore, the Te2-Te3 and Te2-Te4 bond lengths do not significantly change (Fig. 2b). When the doping level reaches to 0.05 $h$/Te (olive line), eigenstates round the G point become unfilled (Fig. 2d). We depicted the wave function norm of the VB1 state at G ($\psi^a_{VB1,G}$) in Fig. 2e, which is a bonding state of the Te2-Te4 and Te4-Te5 bonds. Removing electrons from this state undercuts their bonding and thus leads to elongated bond lengths; this explains the largely elongated covalent bonds (Te4-Te2/Te5) and shortened CLQB bonds (Te3-Te2/Te5) under 0.05 $h$/Te doping (Fig. 2b). Supplementary Figure S7 presents the layer-dependent energy of $\psi^a_{VB1,G}$. Its energy drops from 2L to 3L and 4L, consistent with the enlarged critical doping levels of the $\alpha$-$\beta$ transition, as shown in Fig. 1e.

**Semiconductor to metal transition.** An $\alpha$-bilayer is unstable and is prone to transform into a $\beta$-bilayer upon electron doping. Both $\beta$- and $\gamma$-bilayers share the same feature that they are comprised of rhomboid chains, but in a parallel and a network forms, respectively, as shown in Fig. 1b and 1c. It is thus expectable that the $\gamma$-bilayer (Fig. 3a) shows better stability than the $\beta$-bilayer. A metallic $\delta$-bilayer also looks chain-like but it is comprised of zigzag chains (Fig. 3b). Here, the competition of $\gamma$- and $\delta$-phases is thus highly relevant to the comparison of stability of these chains. We plotted distances $d_1$, $d_2$ and $d_3$ in Fig. 3c that they reflect the inter-chain couplings among rhomboid or zigzag chains. In terms of the $\gamma$-bilayer, either the layer thickness ($d_1$, blue squares) or the interlayer distance ($d_2$, blue circles) slowly shrinks with respect to the concentration of doped electrons; this is relevant with the lowest conduction band (CB1) states around the G point ($\psi^\gamma_{CB1,G}$), as indicated in Fig. 3d and 3e. State $\psi^\gamma_{CB1,G}$, an inter-sublayer bonding state (Fig. 3f), becomes occupied upon electron doping, which strengthens the interactions between the middle and top/bottom sublayers and thus reduce the layer thickness. The reinforced inter-sublayer interaction thus keeps the in-plane inter-chain distance ($d_3$, blue pentagrams) nearly unchanged up to a level of 0.14 $e$/Te and slightly increased beyond that level due to a finite Poisson's ratio. These tendencies of bond-length variations suggest that the rhomboid chains are still network-like, in different from isolated zigzag chains as δ-bilayer behaves under high doping levels.

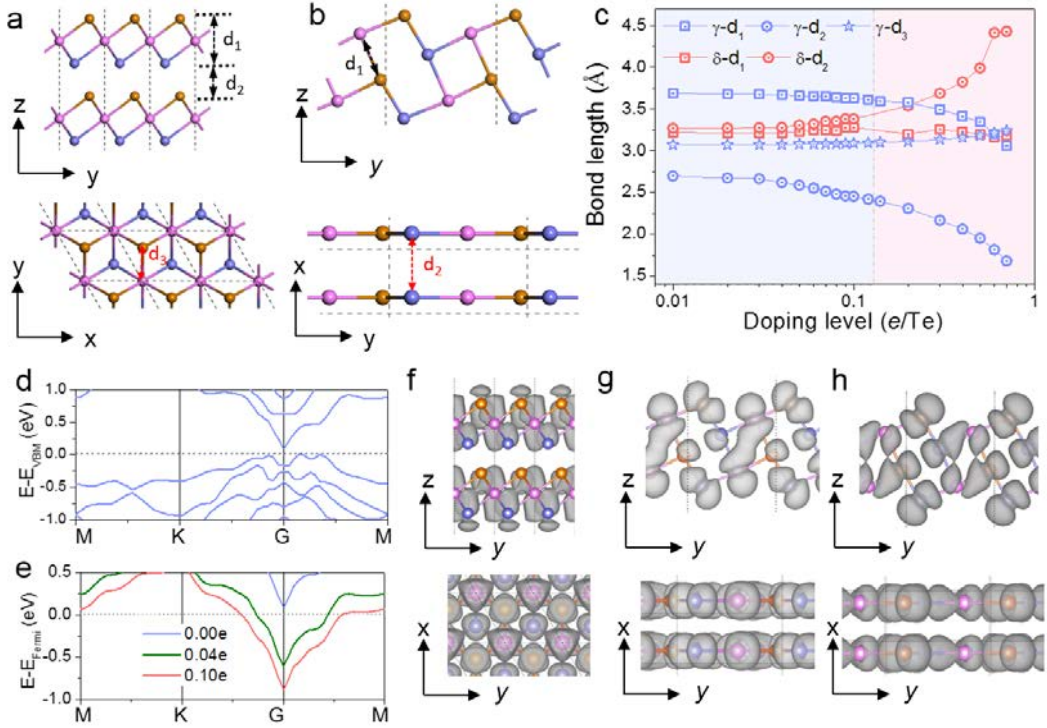

**Fig. 3** Details of the γ-δ phase tansition in a Te bilayer. **a-b** Side- and Top- views of γ- and δ-Te bilayers. The inter- and intra-layer chain-chain distances and layer thickness are marked with dashed arrows. **b** The distances marked in **a-b** as a function of doping level. The blue region represents the energetically stable region of γ phase and pink region for the δ phase. **d** Electronic band structure of bilayer Te in γ phase. **e** The evolution of the bandstructures of the highest valence band (VB) under different charge doping level. (ff) Side- and top-views of the wavefunction norm of the CB state of γ phase at G using an isosurface of 0.001e Bohr$^{-3}$. **g-h** Side- and top-views of the wavefunction norm of the CB state of δ phase at S and X, respectively.

The δ-phase coexists with the γ-phase in all considered doping levels. Its CB1 states around the S ($\psi^{\delta}_{CB1,S}$) and X ($\psi^{\delta}_{CB1,X}$) points are two anti-bonding states (Fig. 3g and 3h) and are filled under electron doping (see Supplementary Figure S8); this gives rise to continuously elongated intralayer distance $d_2$ (red circles). In terms of the interlayer region, $\psi^{\delta}_{CB1,S}$ is a bonding state but $\psi^{\delta}_{CB1,X}$ is an antibonding one. Distance $d_1$ is, therefore, nearly unchanged until the doping level reaches 0.8 $e$/Te. The $\delta$-bilayers can be thus regarded as gradually isolated into six-Te-four-Te chains formed by interlayer bonding. However, $\gamma$-bilayers retain a highly distorted rhomboid network where Te-Te bonds are bent, stretched and compressed under doping due to structural constrains in 2D. The constrain in six-four chains in $\delta$-phase is, however, relaxed in the $x$ direction. Therefore, the strain-relaxed six-four chains are energetically more favored under the doping induced strain. These results also suggest that electron doping might be a route of synthesizing $\delta$-Te nanoribbons.

**Surface chirality**. Surfaces of Te few-layers, with broken translational symmetry and unbalanced interlayer coupling, may play a role in phase transitions. A Te-trilayer is the thinnest layer being comprised of both surface and inner layers. A chiral-$\alpha$+$\delta$ phase emerges in Te-trilayers. Fig. 4a plots its structure that contains one inner $\delta$- and two surface chiral-$\alpha$-chains, which could be directly transformed from an $\alpha$-trilayer (Fig. 4b). The transition brings central-reversion symmetry and two surface layers with opposite chirality to the new phase. The transition occurs in a range from 0.05 $e$/Te (energetic criterion) to 0.09 $e$/Te (structural criterion), see Fig. 1e. This chiral-$\alpha$+$\delta$ phase persists up to a doping level of 0.30 $e$/Te (Fig. 1e) and then transforms into pure $\delta$-phase (Fig. 4c) upon additional doping. We plotted eight bond lengths in Fig. 4d. The side-view (x-z plane) of the neutral structure shows distorted triangles in the surface $\alpha$-chains where surface bond lengths Te1-Te2 (unfilled red squares) and Te10-Te13 (unfilled green pentagrams) are smaller than those of other bonds. These two shortened bonds result in a central inversion symmetry, introducing chirality to both wavefunctions of CB1 and doping induced charge reduction (Supplementary Figure S9).

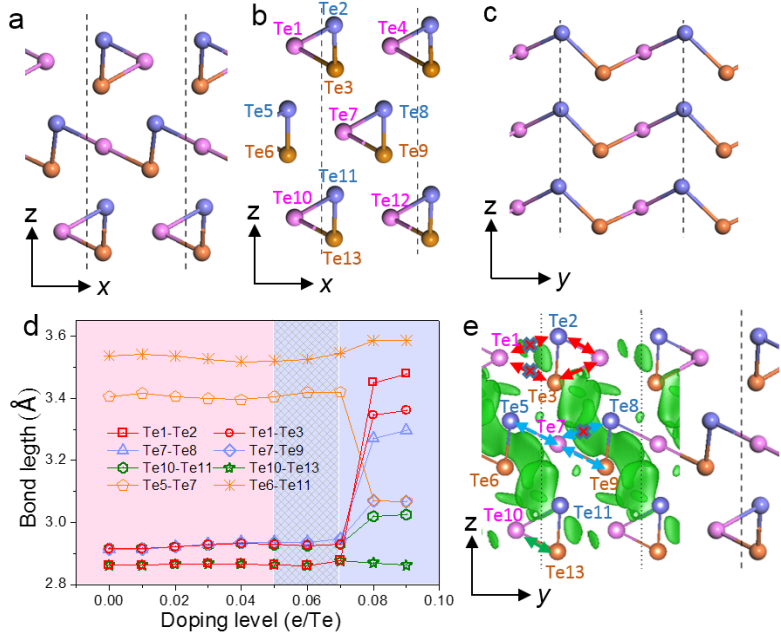

**Fig. 4** Details of phase transition from α- to chiral-α+δ Te-tri-layers. Side views of Te trilayers in the **a** chiral-α+δ-, **b** left-α- **b** and **c** γ-phases, respectively. **d** Distances marked in **b** as a function of electron doping level. The pink region represents the stable region of the α phase and the blue region for the chiral-α+δ phase. The meshed part indicates the transition region between the α and chiral-α+δ phases. **e** Differential charge density (DCD) between a 0.09 $e$/Te doped and a neutral chiral-α+δ-Te tri-layer plotted with an isosurface value of 0.0001$e$/Bohr$^3$. Green isosurface contours indicate charge depletion after electron doping. The breakdown or formation of covalent bonds before and after the phase transition are marked with arrows and crosses.

Fig. 4e maps the charge reduction compared between doping levels of 0.09 $e$/Te and 0.00 $e$/Te on a chiral-α+δ structure. The reduction explicitly shows chirality on the both surface α-chains. In the bottom layer, the reduction is mainly found between in-plane α-chains, resulting in the lattice slightly expanded and the triangle appreciably rotated (3.18°), consistent with the nearly unchanged Te10-Te13 and slightly elongated Te10-Te11 shown in Fig. 4d. Pronounced charge reduction was found around bonds Te6-Te11 and Te3-Te8, which suggests applicable interatomic repulsion upon electron doping, leading to elongated Te6-Te11 (Te3-Te8) (Fig. 4d). In addition, charge reduction also weakens bonds Te7-Te8 (from 2.95Å to 3.27Å) and moves

Te7 towards Te5 and Te6, leading to a reinforced Te5-Te7 bond (from 3.42 Å to 3.07 Å). As a result of both effects, the originally non-tilted Te8-Te9 (Te5-Te6) and 1.7° anti-clockwise-tilted Te2-Te3 become 6.5°/3.2° clockwise-titled and the middle α-layer transforms into a δ-layer. For the top-layer, the charge reduction undercuts both Te1-Te2 and Te1-Te3 bonds (from 2.92/2.86 Å to 3.48/3.36Å, respectively), which substantially pushes Te4 moving leftward forming two new covalent bonds with Te2 and Te3. This movement transforms *l-α-* to *r-α-* chains.

**Discussion**

Raman spectroscopy is a feasible technique for experimental identification of these phases. Fig. 5a show the theoretical Raman spectra of the *α*, *β* and *γ* bilayers. They show a few key identifications for distinguishing different phases. In terms of the *β*-phase, mode $A_1^{10}$ seems an ideal indicator for the appearance of the *β*-phase that its frequency is nearly 10 cm$^{-1}$ lower than that of the $A_1^{11}$ mode (*α*- and *β*-phases) and is over 17 cm$^{-1}$ larger than the $A_1^7$ (*α*-phase) or $A_1^8$ (*β*-phase) mode. Another indicator was found for the transition between the *α*- and *γ*-phases, the frequency of a Raman activated mode $A_1^4$ (*γ*-phase) is over 30 cm$^{-1}$ larger than modes $A_1^{12}$ and $A_1^{11}$ (*α*-phase). By these distinct indicators, these three phases can be distinguished by Raman spectroscopy.

These transitions in 2L could be understood from an energy-level point of view. Semiconductors usually response against charge doping, but Te few-layers are exceptions that they promote charge doping. Fig. 5b plots the VBM and CBM positions of *α*-, *β*- and *γ*-bilayers. The *β*- and *γ*-bilayers have the highest VBM (-4.67 eV) and lowest CBM (-4.97 eV), respectively. Such a deep CBM of *γ*-bilayer and high VBM of *β*-bilayer are even comparable with and 0.29 eV higher than the VBM of *α*-bilayer (-4.98 eV), respectively; this explains the reason why a tiny amount of electron-doping leads to the *γ*-phase more energetically stable and hole-doping stabilizes the *β*-phase. In light of this, the gating/doping induced phase transitions appear a new degree of freedom to tune the electrode-channel contact. In addition, the doping induced phase changes and the accompanying shifts of valence and conduction bands allow Te few-layers to be adaptively doped. The more the electrons doped, the faster the conduction band lowers (*α* to *γ* with a drop over 1 eV). This adaptive doping also works for the hole doping. If Te few-layers were used as channel materials in an FET, this adaptive property of Te may substantially change the density of carriers or/and carrier mobility with a rather small additional gate voltage; this may offer a chance to break the thermodynamic limit for the subthreshold swing. Since *α* and *γ* bilayers have comparable energies (energetic difference of 1 meV/Te), it is rather likely to build an *α*-*γ* Te heterojunction. This junction takes a type-III alignment that may utilize an extra-sensitive gate control and giant carrier density tunability.

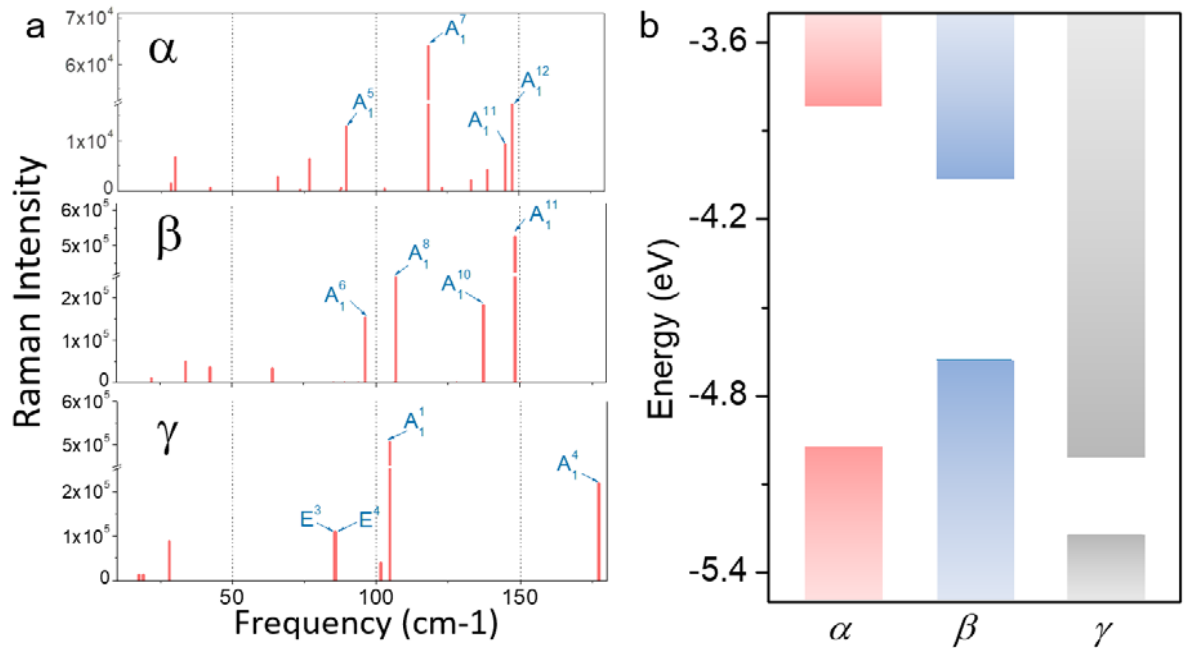

**Fig. 5** Raman intensity of a Te bilayer in different phases. **a** Raman intensities of 2L-Te in *α*, *β* and *γ* phases, respectively. **b** Energy levels of VBM and CBM of bilayer Te in *α*, *β* and *γ* phase. The values were calculated with HSE06+SOC and aligned with the vacuum energies.

**Conclusion**

In summary, we have predicted structural phase transitions among the *α*, *β*, *γ*, *δ* and chiral-*α*+*δ* phases in 2L to 4L Te layers under sole charge doping. Four transitions were discussed in details and understood with doping induced evolution of conduction or valence states and the resulting charge redistributions. The CLQB is peculiarly interesting in Te few-layers that its bond length is usually around 3.4 – 3.5 Å, being only 0.5 – 0.6 Å larger than that of a typical Te-Te covalent bond. Given the only 15% bond length difference and covalent characteristics of Te-Te CLQB and covalent bonds, it is straightforward that these two types of bonding may transform to each other under charge doping, external strain or among the others. Simultaneously, the transformation of bonding thus results in structural phase transitions, accompanied by electric and optical transitions. It suggests that heterojunctions with a type-I, type-II or type-III alignment could be built using only one element by stacking its different allotrope layers together. Another feature of these transitions lies in that a phase may transfer to another phase under working conditions of a device, e.g. new phase induced by laser excited hot electrons or by electric gating resulted charge doping, which may realize more efficient control of its physical properties. We used term ``adaptive doping'' here.

It also suggests that new phases may emerge in nanostructured layered materials because of the surface induced confinement effects. Second harmonic generation is a highly sensitive optical detecting method. Appearances of structural inversion centers in the *β*-, *γ*- and chiral *α*+*δ*-phases should substantially change the signals of it, indicating potential applications of these transitions and the charge stabilized-phases in non-linear optics. Such a phase diversity is, we believe, a unique property of strongly inter-chain coupled 1D-like layered materials. Therefore, we expected other multiple phase transitions in other 1D-like layered materials, like AuSe. These phase transitions also imply high potentials in applications of non-violated memories, actuators and ultra-compact electric devices. In light of this, exploration of different

Te phases may boost an emerging research filed for mono-elemental heterostructures and devices.

**Methods**

Density functional theory calculations were performed using the generalized gradient approximation for the exchange-correlation potential, the projector augmented wave method [41, 42] and a plane-wave basis set as implemented in the Vienna *ab-initio* simulation package (VASP) [43] and Quantum Espresso (QE) [44]. Density functional perturbation theory was employed to calculate phonon-related properties, including Raman intensity (QE), activity (QE) and shifts (VASP), vibrational frequencies at the Gamma point (VASP) and other vibration related properties (VASP). The kinetic energy cut-off for the plane-wave basis set was set to 700 eV for geometric and vibrational properties and 300 eV for electronic structures calculation. A *k*-mesh of 15×11×1 was adopted to sample the first Brillouin zone of the conventional unit cell of few-layer Te in all calculations. The mesh density of *k* points was kept fixed when calculating the properties for bulk Te. In optimizing the system geometry and vibration calculations, van der Waals interactions were considered at the vdW-DF [45, 46] level with the optB88 exchange functional (optB88-vdW) [47-49], which was proved to be accurate in describing the structural properties of layered materials [10, 50-52]. The shape and volume of each supercell were fully optimized and all atoms in the supercell were allowed to relax until the residual force per atom was less than $1\times10^{-4}$ eV·Å$^{-1}$. Electronic bandstructures were calculated using optB88-vdW functional and hybrid functional (HSE06) [53, 54] with and without spin-orbit coupling (SOC). Surface maps of valence and conduction bands shown in Fig. 2e-2l were calculated using the optB88-vdW functional.

Charge doping on Te atoms was realized with the ionic potential method [55], which was used to model the chare transfer from graphite substrates. For electron doping, electrons are removed from a 4d core level of Te and placed into the lowest unoccupied band. For hole doping, electrons was removed from the valence band by adding a negative potential into the 4d core level of those three Te atoms.

## Conflict of interest

The authors declare no conflict of interests.

## Acknowledgments

This project was supported by the National Natural Science Foundation of China (Grant Nos. 11274380, 91433103, 11622437, 61674171 and 61761166009), the Fundamental Research Funds for the Central Universities of China and the Research Funds of Renmin University of China (Grant No. 16XNLQ01), the Research Grant Council of Hong Kong (Grant No. N_PolyU540/17), and the Hong Kong Polytechnic University (Grant Nos. G-SB53). C.W. was supported by the Outstanding Innovative Talents Cultivation Funded Programs 2017 of Renmin University of China. Calculations were performed at the Physics Lab of High-Performance Computing of Renmin University of China and the Shanghai Supercomputer Center.

## Appendix. Supplementary materials

Supplementary materials accompanying this article at.

# Supporting Information

# Charge tunable structural phase transitions in few-layer tellurium

Cong Wang[1], Jingsi Qiao[1], Yuhao Pan[1], Linwei Zhou[1], Xianghua Kong[1], Zhihai Cheng[1], Yang Chai[2] and Wei Ji[1, *]

*

[1]*Beijing Key Laboratory of Optoelectronic Functional Materials & Micro-Nano Devices, Department of Physics, Renmin University of China, Beijing 100872, P. R. China*

[2]*Department of Applied Physics, The Hong Kong Polytechnic University, Hung Hom, Kowloon, Hong Kong, P. R. China*

* wji@ruc.edu.cn

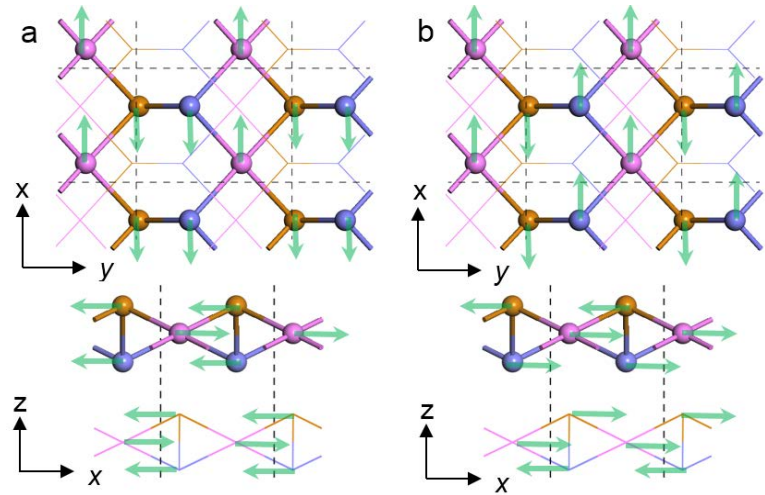

**Supplementary Figure S1. Schematic diagrams of vibrational displacements for imaginary vibration modes of bilayer Te in β phase.** Figure S2a illustrates the vibrational displacement for the imaginary vibration mode of the neutral bilayer β, which indicates that $\beta$-Te is unstable tending to transform to α-Te. Figure S2b shows the vibrational displacement for the imaginary vibration mode appearing with electron doping from 0.04e/Te to 0.1e/Te. Calculations confirmed the stability of l-α+r-α chiral α-Te phase arising from this imaginary vibration mode. The meta-stable l-α+r-α chiral α-Te phase is energetically more stable than other phases except γ phase in a range from 0.04e/Te to 0.1e/Te.

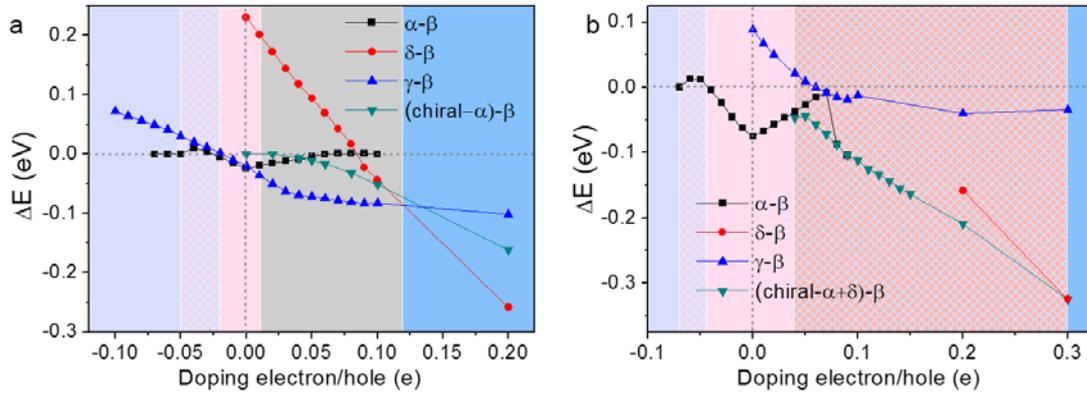

**Supplementary Figure S2. Relative total energies of bi- and tri-layer Te in different phases as a function of electron/hole doping level.** The total energies of the β-bilayer were chosen as the energy reference and the regions with different colors represent the energetically favored regions for each phase. Four phase transitions in 2L and 3L mentioned in the manuscript can be observed here.

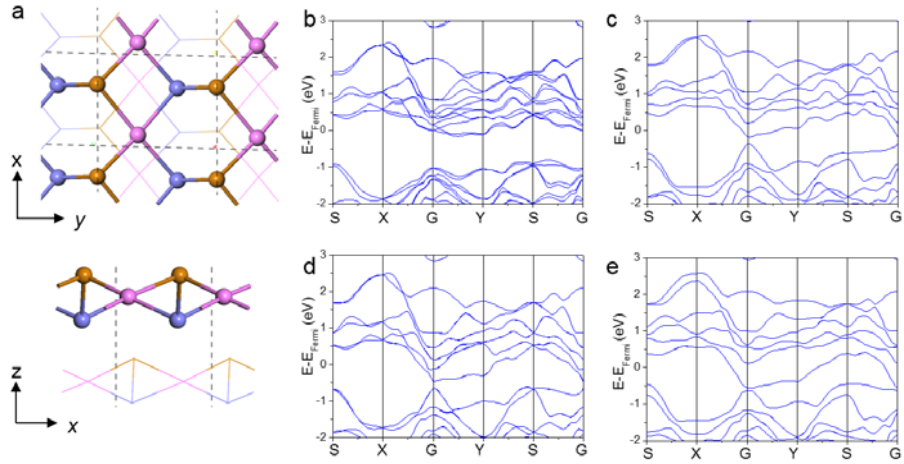

**Supplementary Figure S3. l-α+r-α chiral α-Te in bilayers Te.** (a)Top- and side-view of l-α+r-α chiral α-Te bilayer. Atoms in the top layer were represented by balls while thoese in the bottom layer were shown by lines. (b-e) Electronic band structures of electron doped bilayer Te in different phases: (b) α phase under 0.04e/Te, (c) chiral α phase under 0.04e/Te, (d) β pahse under 0.1e/Te, (e) chiral α phase under 0.1e/Te. Electron doped chiral α phase shows similar band structures to β pahse.

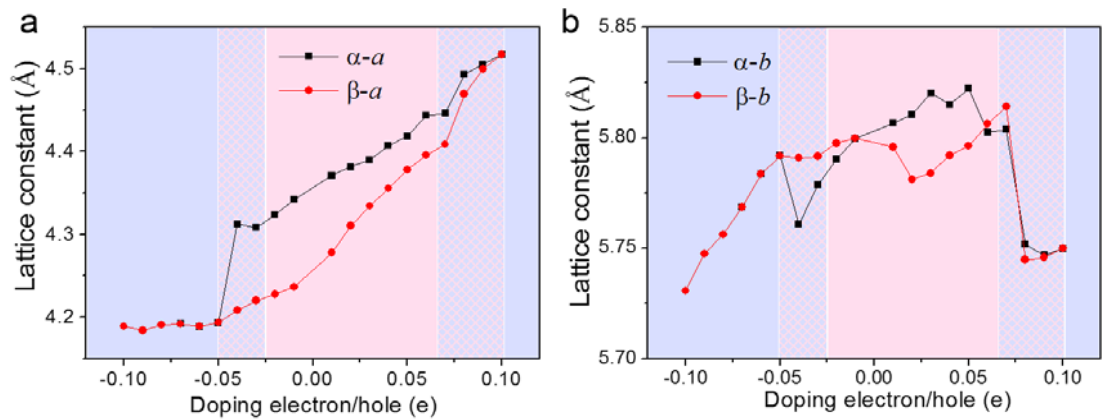

**Supplementary Figure S4. Lattice constants of bilayer Te in α and β phase as a function of electron/hole doping level.** As the increase of the doping level, lattice constant *a* continues expanding, which reflects the strain induce by charge doping. As a result of the varied bond lengths discussed in the manuscript, lattice constant b expands under an electron doping but shrinks in the positively charged region. A slight increase followed with an abrupt drop was found at a hole doping level of 0.04-0.05 h/Te, consistent with the tendency of the bond length changes of Te2-Te3.

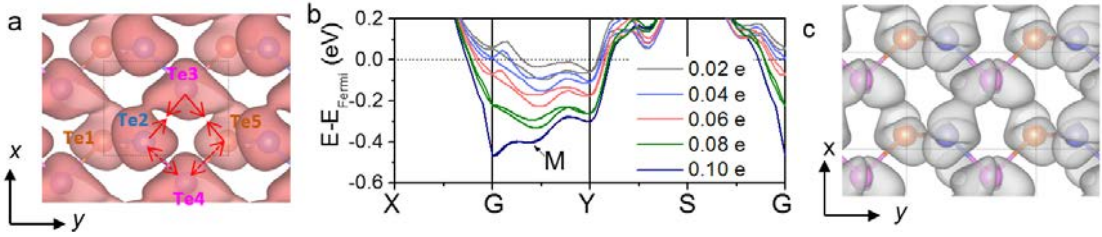

**Supplementary Figure S5. Electron doping induced α-β phase tansition in bilayer Te.** (a) Differential charge density of electron doped bilyer Te-2L-α using an isosurface of 0.0001e Bohr$^{-3}$. (b) The evolution of the bandstructures of the lowest conduction band (CB) under different electron doping level. (c) Top view of the wavefunction norm of the CB state of α phase at G. Figure S3a plots an electron doping DCD by a comparison between the densities of a 0.05 e/Te doped and a neutral alpha-Te layers. It indicates an enhanced inter-chain attraction between Te3 and Te2/Te5, although most exceeding charges are located around Te atoms. The DCD could be well explained from the electronic structural point of view. Figure S3b shows the bandstructures of the lowest conduction band (CB) at electron doping levels of 0.02 e/Te to 0.10 e/Te with a step of 0.02e/Te. It turns out that most CB states between the G and Y points are occupied with electron doping. We plotted the wavefunction norm of the lowest CB at the M point ($\psi^a_{CB,M}$) where the CB has the lowest eigenenergy. Figure S3c explicitly shows that $\psi^a_{CB,M}$ is a bonding state of bonds Te2-Te3 and Te3-Te5 and it is occupied from the 0.02 e/Te doping level. Their bond lengths, therefore, exhibit a gradually shortened tendency rather than an abrupt change with respect to electron doping, consistent with the data presented in Fig. 2b.

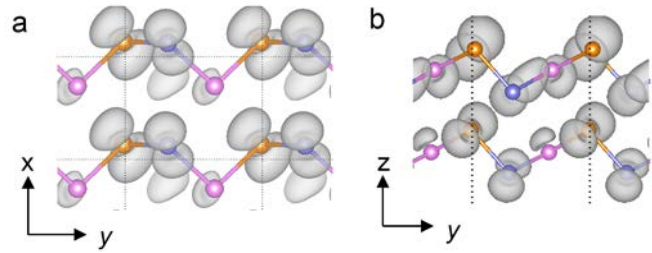

**Supplementary Figure S6. Top and side view of the wavefunction norm of the VB state of α phase at the VBM point.** At a hole doping level up to 0.04 h/Te, eigenstates of VB were unoccupied between the S and G points of the BZ. These states are neither bonding nor anti-boding states. Thus the bond lengths Te2-Te3 and Te2-Te4 change little as hole doping as shown in Figure 3c.

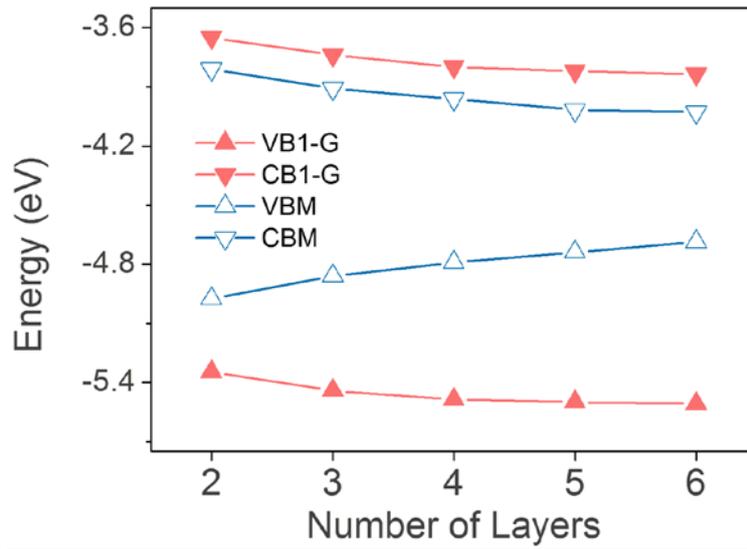

**Supplementary Figure S7. Layer-dependent energy levels of CB1 and VB1 states in few-layer α-Te at G point.** The energy levels of CBM and VBM reported in ref. science bulletin are also shown here. The values are calculated by HSE06 functional inclusion of SOC and aligned with the vacuum energies. The energy of $\psi^{\alpha}_{VB,G}$ drops from 2L to 3L and 4L, consistent with the enlarged critical doping levels of the α-β transition as shown in Fig. 1e.

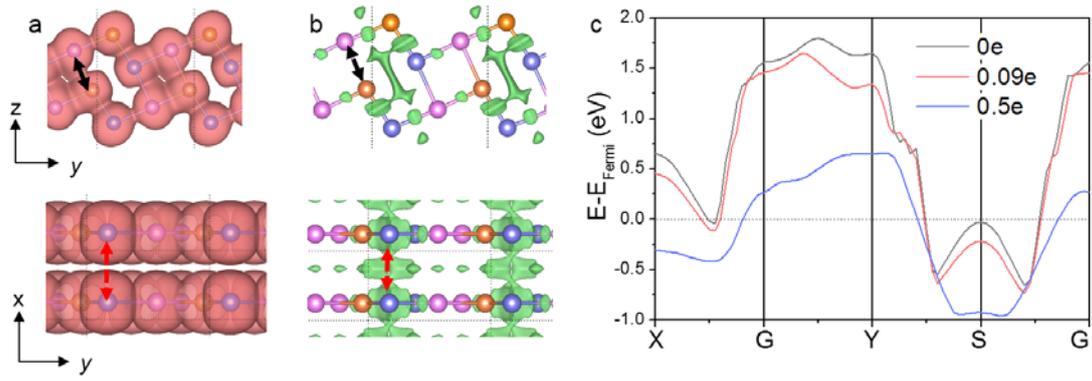

**Supplementary Figure S8. Electronic structure of electron doped bilayer Te in δ phase.** (a-b) Differential charge density of electron doped bilyer Te-2L-δ using an isosurface of 0.0001e Bohr$^{-3}$. The red (a) and green (b) isosufacre correspond to the charge accumulation and reduction after electron doping, respectively. Charge accumulation was found along the δ chain and at the interlayer region, which suggested the enhanced interlayer bonding. Charge reduction mainly occurs as the inter-chain region, suggesting the formation of isolated six-Te-four-Te chains. (c) The evolution of the bandstructures near the fermi level under different electron doping level. The CB states around the S ($\psi^\delta_{CB,S}$) and X ($\psi^\delta_{CB,X}$) points are filled by electron doping.

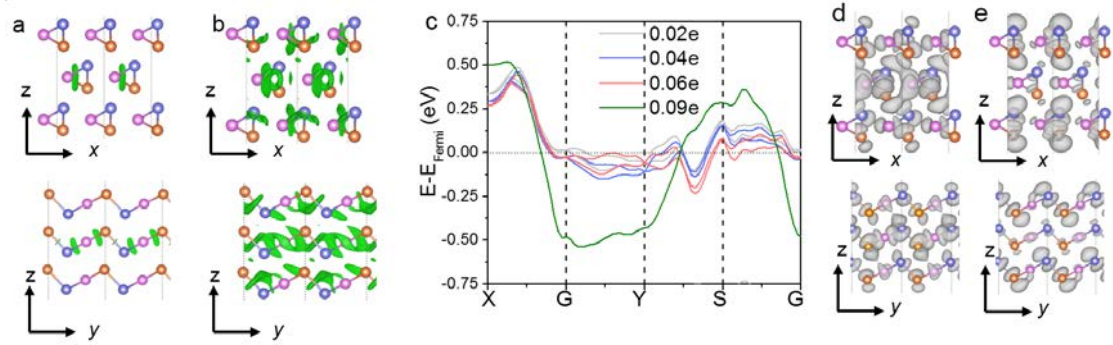

**Supplementary Figure S9. Electronic structure of electron doped trilayer Te.** (a-b) Side and top views of the differential charge density (DCD) of trilayer Te in α phase with electron dopant concentration 0.02e/Te and 0.04e/Te, respectively. The isosurface is set as 0.00004e Bohr$^{-3}$. The green isosurface corresponds to the charge reduction after electron doping. (c) The evolution of the bandstructures of the lowest conduction band (CB) under different electron doping level. The CB states around the G and Y points are filled when the doping electron exceeds 0.09 e/Te. (d-e) Side and top view of the wavefunction norm of the CB state of α phase at the G and S point, respectively. Surfaces in trilayer introduces the emergence of central inversion symmetry and chirality, as discussed in the manuscript. The chirality can be observed in both DCD and the wavefunction norm.

**Table S1**
As each layer may take left-hand α (l-α), right-hand α (l-α) and δ chains, few-layer Te may have considerable arrangement and combinations. The number of possible phases consist of these three kinds of layers are listed as follow. As the number of layer increases, the diversity of phases increases rapidly and reaches 25 for 4L.

| Number of layers | Number of phases |
|---|---|
| 1 | 2 |
| 2 | 4 |
| 3 | 10 |
| 4 | 25 |
| 5 | 70 |
| 6 | 196 |
| 7 | 574 |
| 8 | 1681 |